# Multiple-Modes Scanning Probe Microscopy Characterization of Copper doped Zinc Oxide (ZnO:Cu) Thin Films

Juanxiu Xiao,[1,2] Tun Seng Herng,[3] Jun Ding[3] and Kaiyang Zeng[2,*]

1. State Key Laboratory of Marine Resource Utilization in South China Sea

Hainan University, No. 58, Renmin Avenue, Haikou, Hainan, China, 570228

2. Department of Mechanical Engineering

National University of Singapore, 9 Engineering Drive 1, Singapore, 117576

3. Department of Materials Science and Engineering

National University of Singapore, 9 Engineering Drive 1, Singapore 117576

\* Corresponding author: Prof. K.Y.Zeng, Email: mpezk@nus.edu.sg

Tel: (+65) 6516 6627, Fax: (+65) 6779 1459,

**Abstract**

This paper presents multiple-modes Scanning Probe Microscopy (SPM) studies on characterize resistance switching (RS), polarization rotation (PO) and surface potential changes in copper doped ZnO (ZnO:Cu) thin films. The bipolar RS behavior is confirmed by conductive Atomic Force Microscopy (c-AFM). The PO with almost 180° phase angle is confirmed by using the vertical and lateral Piezoresponse Force Microscopy (PFM). In addition, it elucidates that obvious polarization rotation behavior can be observed in the sample with increasing Cu concentration. Furthermore, correlation of the RS behavior with PO behavior has been studied by performing various mode SPM measurements on the same location. The electric field resulted from the opposite polarization orientation are corresponded to the different resistance states. It is found that the region with the polarization in downward direction has low resistance state (LRS), whereas the region with upward polarization has high resistance state (HRS). In addition, the Piezoresponse Force Spectroscopy (PFS) and Switching Spectroscopy PFM (SS-PFM) measurements further confirm that the existence of the built-in field due to the uncomplemented polarization may affect the depletion region and hence contribute to the RS behavior. In addition, Kelvin Probe Force Microscopy (KPFM) results show that, when ZnO-based thin films is subjected to negative and then followed by positive sample bias, injection charge limit current is dominated.

# 1. Introduction

Resistance random access memories (RRAMs) based on transitional metal oxide (TMO) have enormous potentials in the applications for next generation nonvolatile memory [1-4]. The II-VI compounds, such as ZnO, usually have similar structures to silicon and have advantages of easier fabrication [5]. The resistive switching (RS) behavior in ZnO-based materials have been extensively studied in recent years, especially due to their easier switching properties, high $R_{OFF}/R_{ON}$ ratio, low set and reset voltages, and higher breakdown voltages [6-9]. However, the mechanisms underlying the RS behavior remain unclear, as many internal and external factors may affect these mechanisms. Especially, the coupling between the RS behavior and the polarization rotation (PO) behavior was observed in undoped ZnO thin films and nanostructures [9-11]. The PO was called *polarization switching* or *polarization orientation* in the earlier papers; we define it as *polarization rotation* (PO) in this paper, as ZnO based materials are well-known piezoelectric materials, but not conventional ferroelectric materials. However, recent studies have found that, if ZnO based materials are switched to high resistant state (HRS), the ZnO based materials can show the PO phenomena under an external electric field [9, 11-13]. It was also approved that such behavior was intrinsic properties of the ZnO and not electrostatic effects at the sample surface during the measurements [11, 13]. This coupling has potential to make multiple-states of data storage, such as low and high resistance states, and/or up and down polarization states. On the other hand, this PO behavior may also complicate the underlying electronic transport phenomena in those materials. It was reported that the polarization switching in some ferroelectric materials might change the conduction band profile in the contact of the electrode/specimen and result in RS [13-15]. At the same time, the effects of underlying electronic transport and RS behavior on the PO behavior of ZnO-based materials are also barely studied. It was also reported that the polarization might be incompletely screened by free charges and this might result in a built-in field, $E_{built-in}$, which could also contribute to depletion region in the metal/ferroelectric film interface and influenced the polarization switching behavior [16-19]. Hence, it is necessary to investigate the coupling between the RS behavior and the PO behavior in ZnO based thin films.

In addition, the inevitable existence of intrinsic defects and intentional doping may contribute to the underlying mechanisms. Oxygen vacancies, $V_O$, are inevitable to form under oxygen deficient conditions during the film deposition, especially for the Pulsed Laser Deposition (PLD) [9, 20]. Recent studies have indicated that the electrochemical migration of oxygen vacancies in the vicinity of the interface may result in RS and also favorable to the PO behavior [9, 21]. On the other hand, dopants can also tune the band structure of the ZnO samples [22, 23]. Among doped ZnO compounds, the study shows that the resistivity increases exponentially with increasing the concentration of doped copper [12]. It was reported that the PO behavior in copper doped ZnO materials could be also enhanced due to the cation radius difference [12]. Hence, Cu doped ZnO (ZnO:Cu) thin films can be a good system for comprehensive understanding of the coupling of the local RS, the PO and the underlying electronic transport. Therefore, in this study, the effects of Cu concentration and deposition oxygen partial pressure ($P_{O_2}$) will be studied.

To understand the atomistic mechanisms of RS and PO behaviors in the ZnO:Cu thin films, a study at a single defect level is required. The applications of various Scanning Probe Microscopy (SPM) techniques have offered a pathway to address structure, electrical and electromechanical properties on the nanometer and atomic scales in the last three decades [24]. For example, conductive Atomic Force Microscopy (c-AFM) and Piezoresponse Force Microscopy (PFM) have been used extensively to study RS and polarization switching behaviors in various piezo/ferroelectric materials [9, 25-27]. It is also noticed that, by investigating on the surface potential with Kelvin Probe Force Microscopy (KPFM) technique, the migration of charge carriers and polarization bound charges integrated with the underlying RS and polarization switching behaviors can also be distinguished [28, 29]. In the previous studies, we have applied individual SPM based techniques to characterize the various functionalities of undoped and doped ZnO films. For example, PFM has been used to characterize the polarization rotation in ZnO:Cu films [30], whereas c-AFM technique was used to characterize the resistance switching of ZnO:Cu films [31]. Furthermore, PFM and c-AFM techniques were used together to characterize the polarization rotation and resistance switching of the undoped ZnO films [13]. In this study, we will further apply the multiple-modes SPM techniques at the same location to have detail study of the relationships among the resistive switching, polarization re-orientation and surface potential of the ZnO:Cu thin films.

## 2. Materials and Experiments

*2.1 Samples preparation*

The ZnO:Cu samples (with 2 at.% and 8 at.% Cu content) were deposited by using pulse laser deposition (PLD) technique under oxygen partial pressure of $1\times10^{-6}$ Torr and $2\times10^{-4}$ Torr. The thickness of the films was about 240 nm. The used substrates were commercial Pt-coated Si wafer (Addition Engineering Ltd, CA, USA), in which the Pt coating layer acted as bottom electrode for all of the SPM measurements. The details of preparation procedure was described in the previous work [12].

*2.2 Characterization*

The crystallinity of the films was characterized by using X-ray diffraction (XRD, Briker AXS D8 Advance) [12, 32]. It has confirmed that the films are fully crystalline nature with a preferentially [0001] orientation texture. The Cu percentage of the films was investigated by energy dispersive spectroscopy (EDS) and X-ray photoemission spectroscopy (XPS) [12]; and the oxygen vacancy concentration was characterized by performing the Raman spectra and X-ray Absorption Spectroscopy (XAS) analysis [31]. The detail of those measurements will not be repeated here.

To study the correlations between the RS, PO and surface potential in ZnO:Cu samples, two sequence poling processes were first conducted by c-AFM and followed by KPFM and PFM measurements. It should be noticed that, in the c-AFM measurements, the bias was applied through the bottom electrode and the tip was grounded, and hence it is referred as "sample bias" [Figure S1(a), Supplementary

Information]; whereas in the PFM and KPFM measurements, the bias was applied to the conductive tip, and the bottom Pt electrode of the sample was grounded and this is referred as "tip bias" [Figure S1(b), Supplementary Information]. The two sequence poling processes were defined as "box-in-box" and "up-down" poling processes in this paper [Figure S1(c) and S1(d): Supplementary Information]. In the "box-in-box" poling processes, a sample bias of -10 V was firstly applied on a 5×5 µm$^2$ area and followed by a sample bias of 10 V on a 2.5×2.5 µm$^2$ area in the middle of the area poled by -10 V [Figure S1(c), Supplementary Information]. In the "up-down" poling process, sample biases of 10 V, 0 V, -10 V were applied to the three adjacent 3×1 µm$^2$ areas [Figure S1(d), Supplementary Information]. In these poling processes, the scan rate was set as 0.3 Hz. After these poling processes, current image of a larger area (10×10 µm$^2$) that included these biased regions was obtained by c-AFM with a small DC sample bias of 1 V. To study the correlations between the resistance state, polarization orientation and surface potential, the c-AFM biased region was then scanned under KPFM as well as vertical and lateral PFM (V-PFM and L-PFM) modes respectively. In the PFM measurements, 1 V AC tip bias was used; the parameters for KPFM measurement were the same as these used in the previous studies [32-34].

These correlations were further investigated by Piezoresponse Force Spectroscopy (PFS) technique. In the PFS measurement, the tip was fixed at an arbitrary location. A triangular waveform, which was composed by a sequence of square waves supposed with an AC bias (Figure S2 Supplementary Information), was applied to the tip at a frequency of 200 mHz with alternative of DC bias-on and off. In this waveform, $\tau_1$ is the time of bias-off, and $\tau_2$ is the time of bias-on. Both times were set to be 25 ms. To avoid the contribution of electrostatic effect, the bias-induced remnant piezoresponse was acquired at every pulse between the adjacent voltage steps (i.e., bias-off state). The piezoelectric response hysteresis loop, PR, is calculated using the equation PR = A*cos(φ) and usually plotted as a function of applied biases in the shape of the hysteresis loop, where A is the amplitude, and φ is the phase angle, both are obtained at bias off state [35]. With the use of the high-voltage mode, PFS measurement can be performed at the bias higher than ±10 V. Hence, in this study, the pulse voltage was gradually increased until the saturated hysteresis loops were obtained, i.e., the phase angle change is ~180°. Eventually, the biases of ±10 V and ±15 V was applied to the ZnO:Cu (2 at.% and 8 at.%, deposited at $P_{O_2}$=1×10$^{-6}$ Torr) samples, and ±25 V was applied to the ZnO:Cu (2% at.% and 8 at.%, deposited at $P_{O_2}$=2×10$^{-4}$ Torr) samples, respectively. Furthermore, the SS-PFM (Switching Spectroscopy PFM) measurements were conducted in a grid of 64 x 64 points within a scanning area of 2×2 μm$^2$. Both amplitude and phase loops were acquired, and then the piezoresponse hysteresis loops were calculated on all of the 64×64 points. The imprint was then determined from each hysteresis loop and then a two-dimensional imprint map was generated for this 2×2 μm$^2$ area.

All of the SPM measurements are conducted with a commercial system (MFP-3D, Asylum Research, CA, USA), with commercially available Pt-coated Si tips (AC240TM, Olympus, Japan). The tip has the nominal spring constant of 2 N/m and nominal tip radius of approximately 28 nm and average resonance frequency of 65 kHz (specifications from manufacturer).

## 3. Results and discussion

*3.1 Characteristics by Multiple-Modes SPM techniques*

First, "set" and "reset" experiments at a fixed area are conducted with c-AFM technique. For the ZnO:Cu samples deposited under oxygen partial pressure of $P_{O_2}$ = 2×10$^{-4}$ Torr, the current is negligible from the c-AFM measurements (within applied bias of ±10 V) due to the high resistances of the ZnO:Cu samples. However, the ZnO:Cu samples deposited at oxygen partial pressure of $P_{O_2}$ = 1×10$^{-6}$ Torr have shown moderate conductivity. Figure 1(a) and Figure 2(a) are the current images after the "box-in-box" poling processes [defined in Figure S1(c), Supplementary Information] on ZnO:Cu samples (2 at.% and 8 at.%, $P_{O_2}$ = 1×10$^{-6}$ Torr), respectively. In addition, it is found that 10 V can "set" samples from high resistance state (HRS) to low resistance state (LRS), and -10 V can "reset" samples from LRS to HRS. It is also noticed that the magnitude of the current at LRS decreases as the copper concentration increases from 2 at.% to 8 at.%, indicating the sample with 8 at.% Cu has higher resistance. Afterward, the relationships among the RS and the surface potential distribution, as well as the changes of the polarization orientation after the "set" and "reset" experiments in the ZnO:Cu samples are further studied the KPFM, V-PFM and L-PFM measurements. Figures 1(b)-(f) and 2(b)-(f) show the corresponding KPFM, V-PFM and L-PFM images for ZnO:Cu (2 at.%) and ZnO:Cu (8 at.%) sample respectively. The surface potential value of the biased regions is flattened by the potential of the unbiased region as in the previous studies [29-34]. Figures 1(b) and 2(b) are the surface potential images after two times grounded tip scanning. It is found that, comparing with that of the unbiased region, the area biased by -10 V sample bias shows higher surface potential whereas the area biased by 10 V sample bias shows lower surface potential. It is also noted that the surface potential in the -10 V poled area in ZnO:Cu (8 at.%) sample is actually higher than that of the ZnO:Cu (2 at.%) sample after the same poling processes. These results suggest that the ZnO:Cu (8 at.%) sample (deposited at $P_{O_2}$ = 1×10$^{-6}$ Torr) may have a better charge storage behavior, which is consistent with that reported in the previous study [34]. Figures 1(c) and 2(c) are the V-PFM phase images of the ZnO:Cu (2 at.% and 8 at.%) samples. From the histogram of phase images in Figure 2(g), polarizations with phase angle contrast of around 180° can be observed. It is found that numbers of polarizations are at the phase angle of about -45°, and other polarizations are around at the phase angle of about 135°. It is obvious that most of polarizations in the region poled by 10 V (green square: 2.5×2.5 µm$^2$) are switched to upward direction (toward to the sample surface), and at the same time, this region is in HRS. On the other hand, it is also noticed that for the region biased with -10 V (red square: 5×5 µm$^2$), most of polarizations are in downward direction (toward to the bottom electrode), and the region is in LRS. This suggests that the LRS region is associated with the polarizations in the downward direction. There are more polarizations at the phase angle of about -45° in the ZnO:Cu (8 at.%) sample, indicating more polarizations are in the downward direction in this sample. On the other hand, the L-PFM phase image shows a change in the region poled by 10 V [Figures 1(e) and 2(e)], indicating that certain degree of polarization rotation also occurs in the in-plane direction. Furthermore, the corresponding responses can be also observed from the V-PFM amplitude images [Figures 1(d) and 2(d)] and L-PFM amplitude images [Figures 1(f) and 2(f)]. Comparing the multiple-modes SPM results, it can be concluded that, when the sample is subjected to negative sample bias, the emerging of current is observed, which can compensate the downward polarization [Figure 2(g)], and this may favor the incurring current. Therefore, in this negative sample biased region, a surface potential increase is also observed, and this corresponds to the presence of additional positive charges attributing to hole injection. On the other hand, when sample is subjected to positive sample bias, there is no observable current and the polarization is re-orientated to upward direction [Figure 2(g)], the decrease in surface potential should due to compensated negative charge ions diffusion to the sample surface. It should be noticed that as the doped copper concentration increasing, polarization is rotated to downward direction within the region with LRS, and this is significantly different from that in the undoped ZnO film, in which the polarization rotation was only observed in the region with HRS [9].

To further confirm the these results, the correlations among the RS, PO and surface potential in ZnO:Cu (2 at.%) sample are further studied by conducting the "up-down" poling processes [Figure S1(d), Supplementary Information]. The current image shows that the area poled by -10 V bias is in LRS, whereas the area poled by 10 V is in HRS after removing the external electric field [Figure 3(a)]. The changes in the surface potential are also consistent with these observed from the "box-in-box" poling processes. Furthermore, Figure 3(b) shows the surface potential can be easily removed by grounded tip scanning, indicating the charges in the LRS region are mainly screen charges. There is still small numbers of polarizations in the LRS region have re-orientated to the downward direction [Figures 3(c)]. This can be clearly seen from Figure 4, which shows the histogram plots of the phase image [Figure 3(c)] from the areas poled by 10 V and -10 V as well as the unbiased areas, respectively. It shows that the numbers of phase angle higher than 150º (upward polarizations) is slight more than the numbers of phase angle around 0º (downward polarizations) in the area poled by 10 V sample bias [Figure 4(a)]. The peak for downward polarizations (phase angle about 0º) in the area poled by -10 V sample bias (LRS) is very sharp, whereas the peak for upward polarizations (phase angle about 150º) was almost disappeared [Figure 4(b)]; therefore it suggests almost all polarizations in upward direction are switched to downward direction. However, it is also found that the switched polarizations cannot be maintained for long time in the LRS, the polarizations are switched back after about 14 hours [Figure 3(d)].

*3.2 Mechanism of RS behavior*

It was assumed that the RS behavior in the ZnO:Cu films may be dominated by Schottky emissions [12, 32]. This can be explained as following: First, the Pt/ZnO:Cu interface is the Schottky contact because of the differences between the work functions of Pt and ZnO:Cu sample [32]. Next, it is known that the copper ions are occupied on the $Zn^{2+}$ sites [12], which resulted in the formation of the trapping states and shifts the Fermi level toward the valence band, hence reduces the conductivity of the film [35]. In addition, it is also known that oxygen vacancies can act as donors that can release electrons to neutralize the positive charge bound by polarizations or annihilate with holes introduced by the copper ions to bend the conduction band downward [15, 16, 36, 37]. This downward bended band structure can further augment work function difference, which may increase the barrier height of the Schottky contact. Furthermore, the polarization is incompletely screened by free charges [38] and it can result in a built-in field, $E_{bi}$ [16], which may contribute to depletion region in the metal/thin film interface [16, 17]. In order to verify the built-in field effect caused by uncompensated polarization in ZnO:Cu films, PFS experiments are conducted. The positive ($V_p$) and negative ($V_n$) coercive biases can be determined piezoresponse hysteresis loop measured by PFS experiments, hence, some critical parameters can be determined, for example, the imprint bias is defined as $(|V_p| - |V_n|)/2$ and the built-in field is $E_{built-in} = |V_p| - |V_n|$ [36, 39, 40]. Figures 5(a) and 5(b) are the PFS measured phase and amplitude loops for the 2 at.% ZnO:Cu thin film sample deposited at $P_{O_2}$ = 1×10$^{-6}$ Torr. In this measurement, PFS loops are taken from eight random locations on a 2×2 µm$^2$ area of each sample. The calculated piezoresponse hysteresis loop is shown in Figure 5(c). It is clear that phase angles in the ZnO:Cu (2 at.%) sample change approximately 180º as the function of the applied voltage, and this PO phenomena may be attributed by two factors: (i) the partial substitution of host Zn$^{2+}$ ions by smaller Cu$^{2+}$ ions [12]; and (ii) the existence of oxygen vacancy [12, 31, 41]. Furthermore, it is worth to mention that there are two obvious asymmetries in the hysteresis loops [Figures 5(a) and 5(b)]. It is found that the positive coercive bias is 5.76 V and the negative coercive bias is -3.76 V, and this leads to a shift of 2.16 V along the bias axis. This shift is primarily corresponding to the built-in electric field. In addition, this PO behavior is also observed when the sample is placed in the closed cell filled with flowing synthetic air (<5 ppm water) or flowing argon gas (<0.02 ppm water and <0.01 ppm oxygen) respectively, and these measurements in controlled gas environments can avoid the moisture and oxygen effects from the ambient air condition [Figure 5(d)]. It is noted that, when there is no moisture or oxygen in the environments, both positive and negative coercive biases are increased. In addition, the PR loops have become more symmetric and this indicates that the build-in field has been reduced when there is no moisture or oxygen in the environments. Furthermore, the built-in field, i.e., twice of imprint bias is also confirmed by conducting SS-PFM on the same sample. The imprint image can be determined from each hysteresis loop of 64×64 points on 2×2 $\mu m^2$ area (Figure 6). The imprint map shows that there are many points show larger imprint values (blue points). It should be also noticed that the points in red are the locations with un-saturated hysteresis loops [39]. Moreover, this PO behavior is observed in all of the ZnO:Cu samples. Table 1 summarizes the average built-in fields determined from the SS-PFM results for all of the samples. It is clear that the built-in field increases with the increasing of the copper concentration or the oxygen partial pressure during the film deposition. As the built-in fields are positive, its direction is from the substrate to the SPM tip and this favors the polarization direction of [0001], which is consistent with the polarization oriented in upward direction. As showed in Figures 1(c) and 2(c), the most polarizations in un-biased region has the phase angle of approximately 180º, especially for the ZnO film doped with 8 at.% of Cu, this indicates that the majority of the polarizations are in the upward direction (yellow color); under the -10 V sample bias, the polarizations are changed to approximately 0º (in purple color), and under the 10 V sample bias, the polarization is re-orientated to about 180º again (yellow color). This polarization direction favor was also reported for other materials, such as PZT and BTO [17, 42]. Hence when the external field is applied along the [0001] polarization direction, i.e., under positive sample bias or negative tip bias, there is a forward current (LRS); it is reported the current flow may strongly influence the shape of the piezoresponse loops [43].

## 4. Summary and Conclusion

In this work, the resistive switching behavior is observed in the ZnO:Cu thin films by "set" and "reset" process at a fixed area. With the increasing copper concentration, the film resistance is increased; and HRS and LRS are more distinguished. Increased charge storage and polarization rotation behavior are also observed by defects engineering with doping more copper. By comparing c-AFM, PFM and KPFM results on the same location, clear couplings among RS behavior, PO behavior and surface potential are demonstrated. Finally, the RS behavior affected by the existence of the built-in field is further confirmed by PFS measurements. The results show that both the copper and oxygen vacancy can tune the band structure and depolarization field. These results give a better interpretation of the couplings among the RS behavior, the PO behavior and surface charges distribution in ZnO:Cu thin films; and the exploration of the underlying mechanisms of RS may imply the development of ZnO:Cu as information storage devices.

**Acknowledgments**

This work was supported by Ministry of Education, Singapore, through National University of Singapore (NUS) under the Academic Research Fund (grant R265-000-406-112).

**List of Table**

Table 1: Summary of the average built-in fields determined from the SS-PFM results for each of the samples.

| $P_{O_2}$(Torr) | $1\times10^{-6}$ | $1\times10^{-6}$ | $2\times10^{-4}$ | $2\times10^{-4}$ |
|---|---|---|---|---|
| Cu concentration | 2% | 8% | 2% | 8% |
| $E_{built-in}$(V) | 2.16 | 3.18 | 5.26 | 13.26 |

**Figure Caption**

Figure 1: Correlations between the RS, PO and surface potential in ZnO:Cu (2 at.% sample ($P_{O_2}$ = 1×10$^{-6}$ Torr) by -10 V/10 V ("box-in-box') poling processes. A smaller DC sample bias of 1 V was applied to scan a 10×10 µm² area after applying -10 V on the middle of 5×5 µm² area (red square) and sequentially poling by 10 V on a central 2.5×2.5 µm² area (green square). The scan rate is 1 Hz: (a) current image by c-AFM under sample bias; (b) the corresponding surface potential image by KPFM measurement under tip bias; (c) and (d) the corresponding vertical PFM phase image under tip bias; and (e) and (f) the corresponding lateral PFM phase and amplitude images under tip bias. All of the images are obtained on the same location.

Figure 2: Correlations between the RS, PO and surface charge distribution in ZnO:Cu (8 at.%) sample ($P_{O_2}$ = 1×10$^{-6}$ Torr) by -10 V/10 V ("box-in-box") poling processes. A smaller DC sample bias of 1 V was applied to scan a 10×10 µm² area after applying -10 V on the middle of 5×5 µm² area (red square) and sequentially poling by 10 V on a central 2.5×2.5 µm² area (green square). The scan rate is 1 Hz: (a) current image by c-AFM under sample bias; (b) the corresponding surface potential image by KPFM measurement under tip bias; (c) and (d) the corresponding vertical PFM phase image under tip bias; and (e) and (f) the corresponding lateral PFM phase and amplitude images under tip bias. All of the images are obtained on the same location; (g) the histograms of the 5×5 µm² areas (subjected

to -10 V) in the Figure 3(c) and Figure 4(c), and schematic drawing of the polarization reversal by 10 V/-10 V sample voltage.

Figure 3: Correlations between the RS, PRO and charge distribution in ZnO:Cu (2 at.%) sample ($P_{O_2}$ = 1×10$^{-6}$ Torr) by -10 V/0 V/10 V ("up-down") poling processes. A smaller DC sample bias of 1 V was applied to scan a 6×6 µm$^2$ area after applying -10V/0V/10V on the three adjacent areas of 3×1 µm$^2$ (enclosed by red lines). The scan rate is 1 Hz: (a) current image by c-AFM under sample bias; (b) the corresponding surface potential image by KPFM measurement under tip bias; (c) V-PFM phase image; and (d) V-PFM phase image obtained 14 hours later after the poling process. All of the images are obtained from the same location.

Figure 4: Histograms of the phase image [Figure 3(c)] of (a) the area (3×1 µm$^2$) poled by 10 V; (b) the area (3×1 µm$^2$) poled by -10 V and (c) the unbiased area, respectively.

Figure 5: (a) and (b): PFS measured phase and amplitude loops; (c) calculated piezoresponse hysteresis loops in eight random locations for ZnO:Cu (2 at.%) sample ($P_{O_2}$ = 1×10$^{-6}$ Torr), based on the equation of PR = A*cos(φ), where A is the amplitude at bias off state, φ is the phase angle at bias off state; (d) Hysteresis loops, PR for the same sample by conducting PFS measurements in the ambient air, and in closed cell filled with flowing synthetic air (<5 ppm water) and flowing argon gas (<0.02 ppm water and <0.01 ppm oxygen) respectively.

Figure 6: (a) Imprint image by calculating the values of ( | $V_p$ | - | $V_n$ | )/2 from the SS-PFM hysteresis loops on a grid of 64 x 64 within a 2×2 µm² area from ZnO:Cu (2 at.%) thin film sample ($P_{O_2}$ =1×10⁻⁶ Torr), in which $V_p$ and $V_n$ are the positive and negative coercive biases separately.

Figure 1

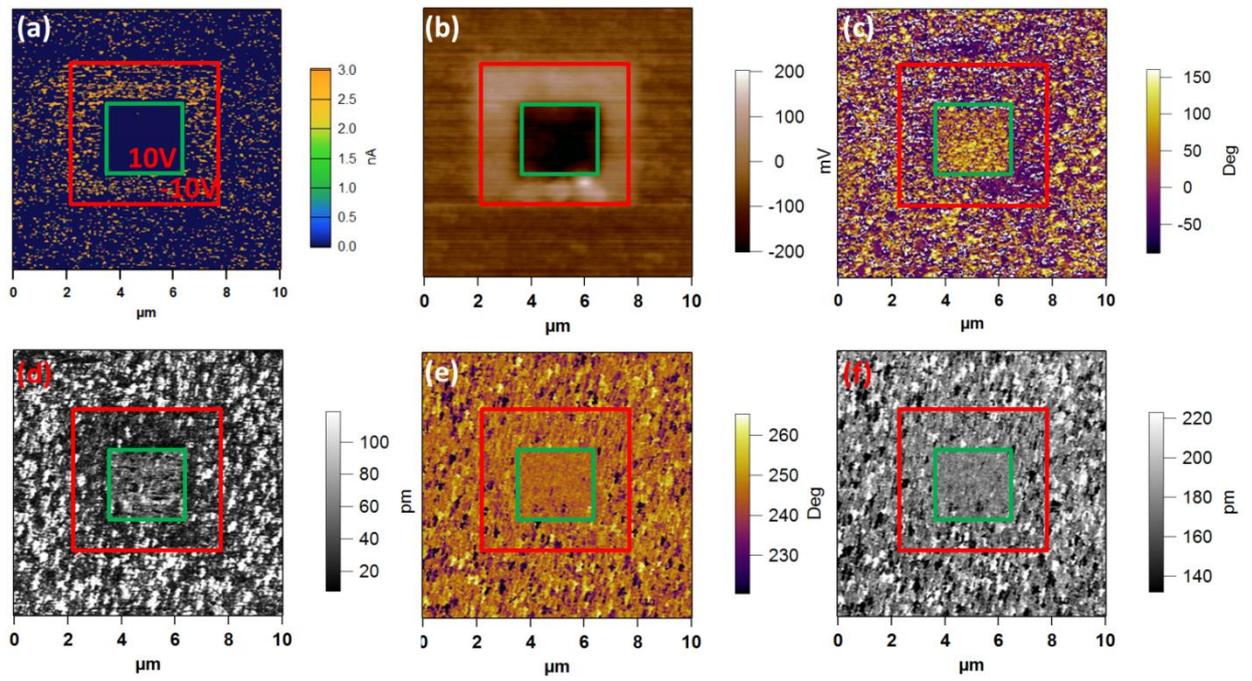

Figure 2

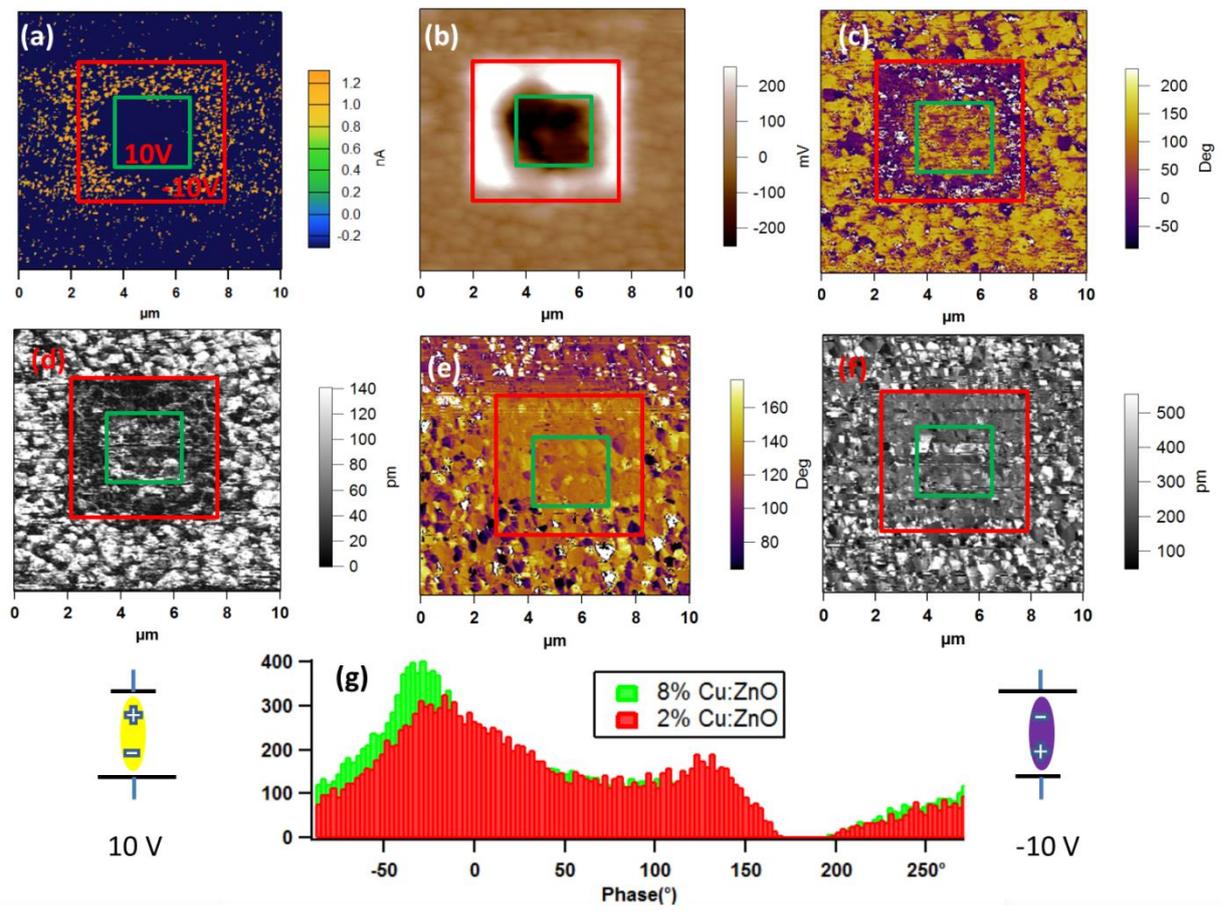

Figure 3

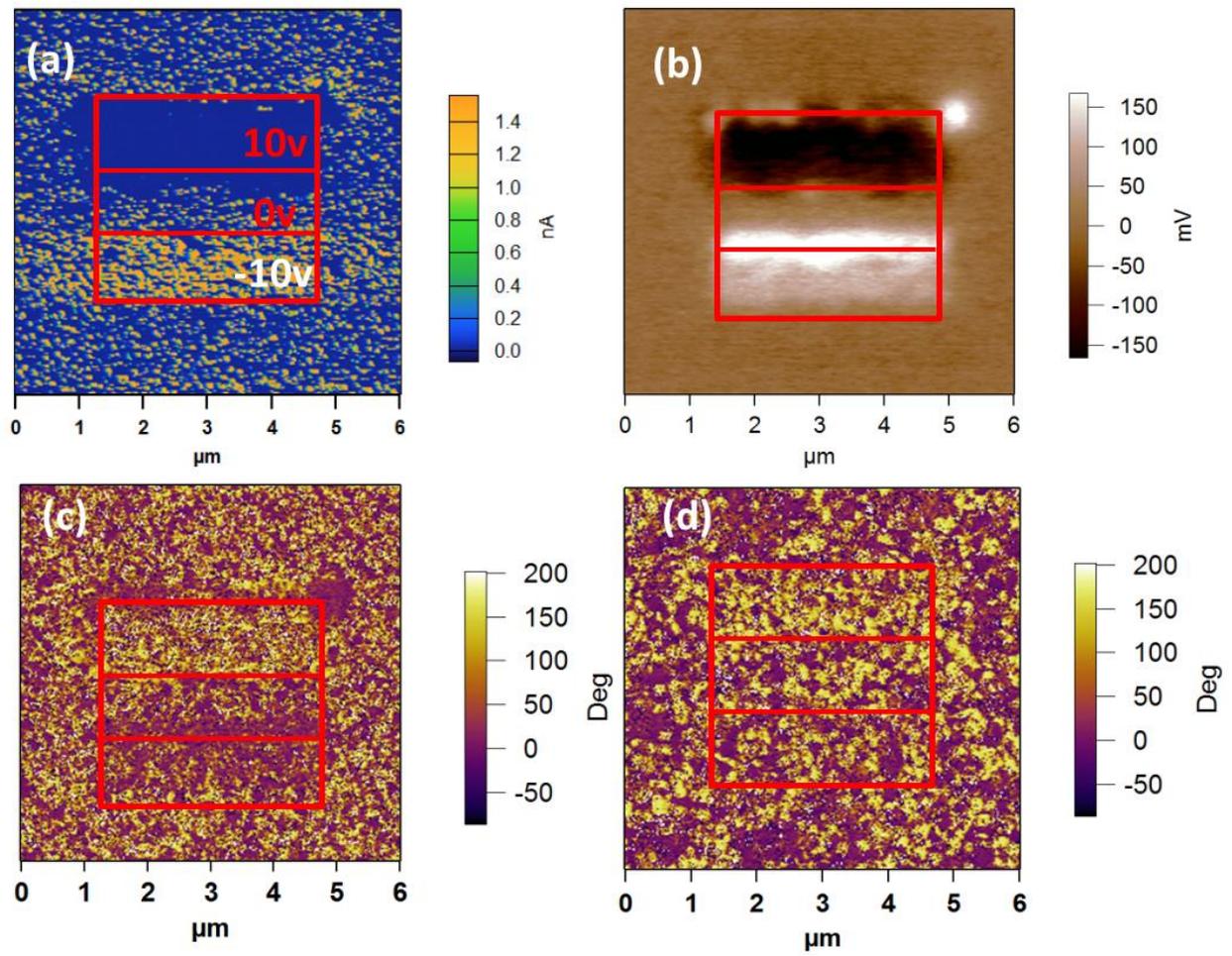

Figure 4

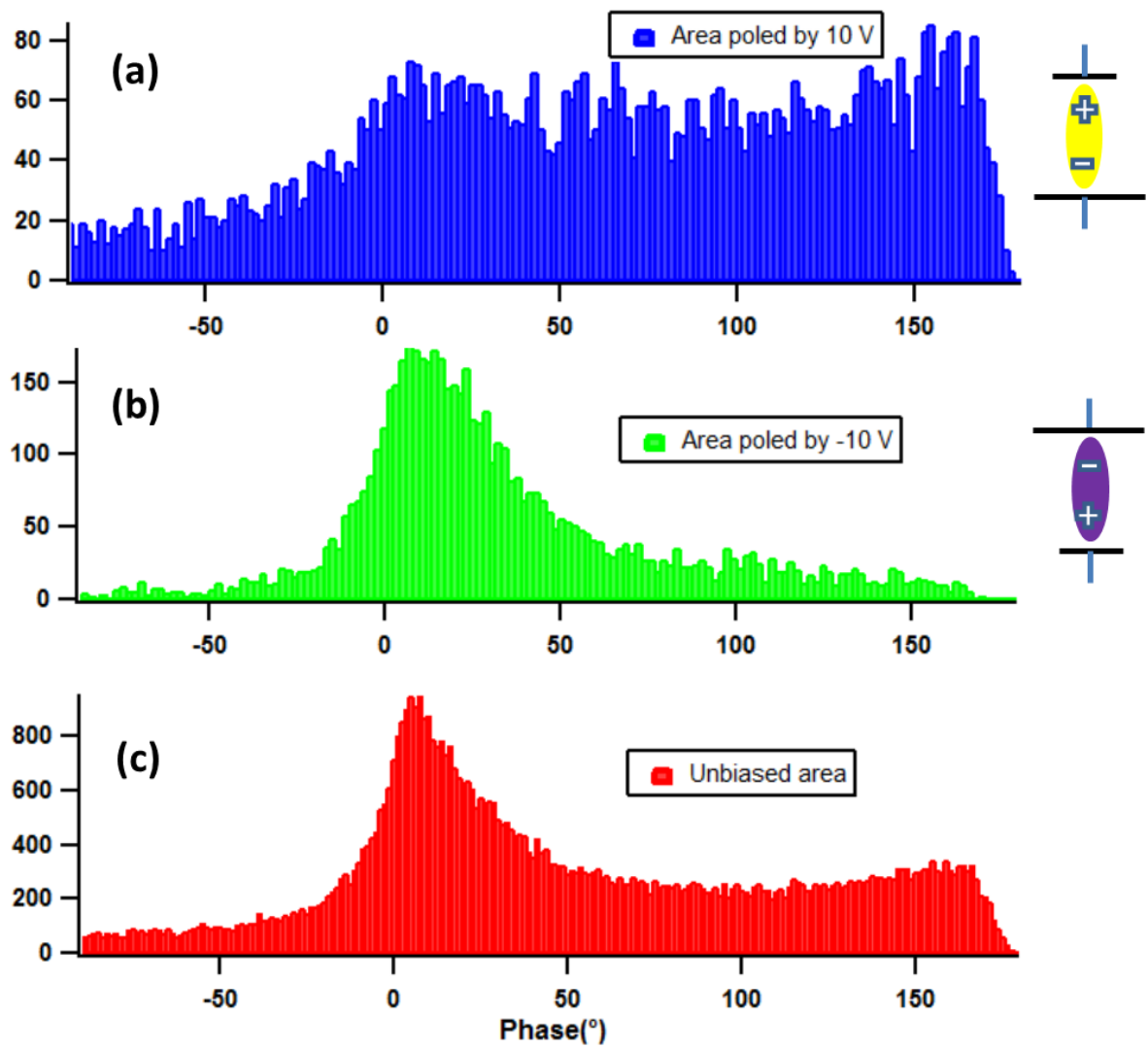

Figure 5

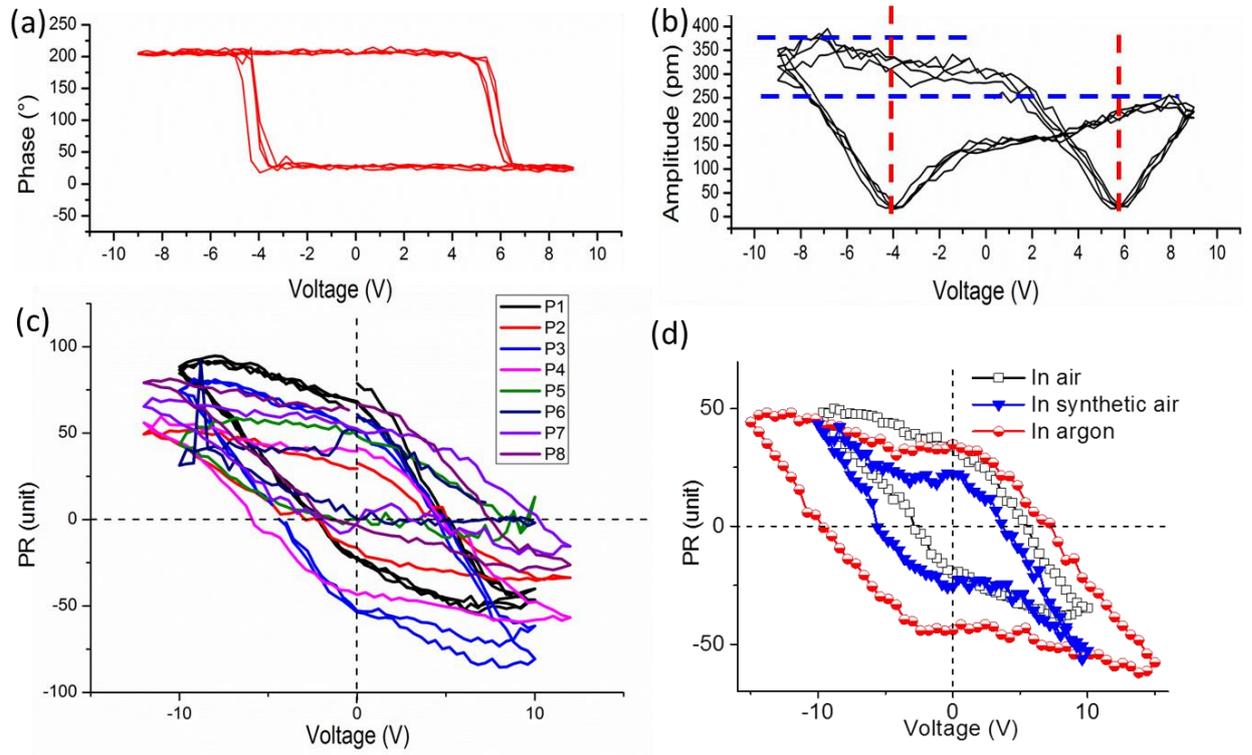

Figure 6

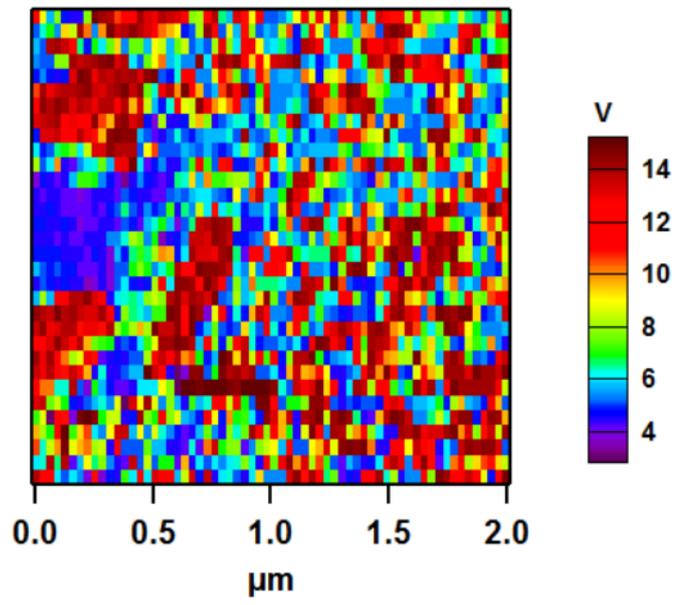